\documentclass[12pt]{article}
\usepackage{epsf,latexsym}
\usepackage{amsfonts,amssymb} 
\epsfverbosetrue
\usepackage[ps,arc,frame]{xy}

\newcommand{\rxyz}[2]{{\begin{xy} 0;<2mm,0mm>:<0mm,2mm>::0;0,
,(5,-2)*{a} ,(10,-2)*{b} ,(15,-2)*{c} ,(2,-5)*{a} ,(2,-10)*{b} ,(2,-15)*{c}
,(5,-5)*\cir(#1,0){} ,(10,-5)*\cir(#1,0){} ,(15,-5)*\cir(#1,0){}
,(5,-10)*\cir(#1,0){} ,(10,-10)*\cir(#1,0){} ,(15,-10)*\cir(#1,0){}
,(5,-15)*\cir(#1,0){} ,(10,-15)*\cir(#1,0){} ,(15,-15)*\cir(#1,0){}
#2\end{xy}}}

\newcommand{\rxyd}[2]{{\begin{xy} 0;<2mm,0mm>:<0mm,2mm>::0;0,
,(5,-2)*{a} ,(10,-1.8)*{\bar{a}} ,(15,-2)*{b} ,(20,-1.8)*{\bar{b}} ,(25,-2)*{c}
,(2,-5)*{a} ,(2,-10)*{\bar{a}} ,(2,-15)*{b} ,(2,-20)*{\bar{b}} ,(2,-25)*{c}
,(5,-5)*\cir(#1,0){} ,(10,-5)*\cir(#1,0){} ,(15,-5)*\cir(#1,0){}
,(20,-5)*\cir(#1,0){} ,(25,-5)*\cir(#1,0){} ,(5,-10)*\cir(#1,0){}
,(10,-10)*\cir(#1,0){} ,(15,-10)*\cir(#1,0){} ,(20,-10)*\cir(#1,0){}
,(25,-10)*\cir(#1,0){} ,(5,-15)*\cir(#1,0){} ,(10,-15)*\cir(#1,0){}
,(15,-15)*\cir(#1,0){} ,(20,-15)*\cir(#1,0){} ,(25,-15)*\cir(#1,0){}
,(5,-20)*\cir(#1,0){} ,(10,-20)*\cir(#1,0){} ,(15,-20)*\cir(#1,0){}
,(20,-20)*\cir(#1,0){} ,(25,-20)*\cir(#1,0){} ,(5,-25)*\cir(#1,0){}
,(10,-25)*\cir(#1,0){} ,(15,-25)*\cir(#1,0){} ,(20,-25)*\cir(#1,0){}
,(25,-25)*\cir(#1,0){} #2\end{xy}}}

\newcommand{\rxydt}[2]{{\begin{xy}
0;<2mm,0mm>:<0mm,2mm>::0;0, ,(5,-2)*{a}
,(10,-1.8)*{\bar{a}}
,(15,-2)*{b}
,(20,-1.8)*{\bar{b}}
,(25,-2)*{c}
,(2,-5)*{a}
,(2,-10)*{\bar{a}}
,(2,-15)*{b}
,(2,-20)*{\bar{b}}
,(2,-25)*{c}
,(5,-5)*\cir(#1,0){}
,(10,-5)*\cir(#1,0){}
,(15,-5)*\cir(#1,0){}
,(20,-5)*\cir(#1,0){}
,(25,-5)*\cir(#1,0){}
,(5,-10)*\cir(#1,0){}
,(10,-10)*\cir(#1,0){}
,(15,-10)*\cir(#1,0){}
,(20,-10)*\cir(#1,0){}
,(25,-10)*\cir(#1,0){}
,(5,-15)*\cir(#1,0){}
,(10,-15)*\cir(#1,0){}
,(15,-15)*\cir(#1,0){}
,(20,-15)*\cir(#1,0){}
,(25,-15)*\cir(#1,0){}
,(5,-20)*\cir(#1,0){}
,(10,-20)*\cir(#1,0){}
,(15,-20)*\cir(#1,0){}
,(20,-20)*\cir(#1,0){}
,(25,-20)*\cir(#1,0){}
,(5,-25)*\cir(#1,0){}
,(10,-25)*\cir(#1,0){}
,(15,-25)*\cir(#1,0){}
,(20,-25)*\cir(#1,0){}
,(25,-25)*\cir(#1,0){}
#2\end{xy}}}

\newcommand{\rxydd}[2]{{\begin{xy} 0;<2mm,0mm>:<0mm,2mm>::0;0,
,(5,-2)*{a}
,(10,-1.8)*{\bar{a}}
,(15,-2)*{b}
,(20,-1.8)*{c}
,(25,-2)*{\bar{c}}
,(2,-5)*{a}
,(2,-10)*{\bar{a}}
,(2,-15)*{b}
,(2,-20)*{c}
,(2,-25)*{\bar{c}}
,(5,-5)*\cir(#1,0){}
,(10,-5)*\cir(#1,0){}
,(15,-5)*\cir(#1,0){}
,(20,-5)*\cir(#1,0){}
,(25,-5)*\cir(#1,0){}
,(5,-10)*\cir(#1,0){}
,(10,-10)*\cir(#1,0){}
,(15,-10)*\cir(#1,0){}
,(20,-10)*\cir(#1,0){}
,(25,-10)*\cir(#1,0){}
,(5,-15)*\cir(#1,0){}
,(10,-15)*\cir(#1,0){}
,(15,-15)*\cir(#1,0){}
,(20,-15)*\cir(#1,0){}
,(25,-15)*\cir(#1,0){}
,(5,-20)*\cir(#1,0){}
,(10,-20)*\cir(#1,0){}
,(15,-20)*\cir(#1,0){}
,(20,-20)*\cir(#1,0){}
,(25,-20)*\cir(#1,0){}
,(5,-25)*\cir(#1,0){}
,(10,-25)*\cir(#1,0){}
,(15,-25)*\cir(#1,0){}
,(20,-25)*\cir(#1,0){}
,(25,-25)*\cir(#1,0){}
#2\end{xy}}}

\newcommand{\rxyddd}[2]{{\begin{xy}
0;<2mm,0mm>:<0mm,2mm>::0;0, ,(5,-2)*{a}
,(10,-1.8)*{\bar{a}}
,(15,-2)*{b}
,(20,-1.8)*{\bar{b}}
,(25,-2)*{c}
,(30,-1.8)*{\bar{c}}
,(2,-5)*{a}
,(2,-10)*{\bar{a}}
,(2,-15)*{b}
,(2,-20)*{\bar{b}}
,(2,-25)*{c}
,(2,-30)*{\bar{c}}
,(5,-5)*\cir(#1,0){}
,(10,-5)*\cir(#1,0){}
,(15,-5)*\cir(#1,0){}
,(20,-5)*\cir(#1,0){}
,(25,-5)*\cir(#1,0){}
,(30,-5)*\cir(#1,0){}
,(5,-10)*\cir(#1,0){}
,(10,-10)*\cir(#1,0){}
,(15,-10)*\cir(#1,0){}
,(20,-10)*\cir(#1,0){}
,(25,-10)*\cir(#1,0){}
,(30,-10)*\cir(#1,0){}
,(5,-15)*\cir(#1,0){}
,(10,-15)*\cir(#1,0){}
,(15,-15)*\cir(#1,0){}
,(20,-15)*\cir(#1,0){}
,(25,-15)*\cir(#1,0){}
,(30,-15)*\cir(#1,0){}
,(5,-20)*\cir(#1,0){}
,(10,-20)*\cir(#1,0){}
,(15,-20)*\cir(#1,0){}
,(20,-20)*\cir(#1,0){}
,(25,-20)*\cir(#1,0){}
,(30,-20)*\cir(#1,0){}
,(5,-25)*\cir(#1,0){}
,(10,-25)*\cir(#1,0){}
,(15,-25)*\cir(#1,0){}
,(20,-25)*\cir(#1,0){}
,(25,-25)*\cir(#1,0){}
,(30,-25)*\cir(#1,0){}
,(5,-30)*\cir(#1,0){}
,(10,-30)*\cir(#1,0){}
,(15,-30)*\cir(#1,0){}
,(20,-30)*\cir(#1,0){}
,(25,-30)*\cir(#1,0){}
,(30,-30)*\cir(#1,0){}
#2\end{xy}}}

\newcommand{\goth}[1]{\mathfrak{#1}}
\newcommand{\double}[1]{\mathbb{#1}}

\newcommand{\cc}{\double{C}}

\newcommand{\rr}{\double{R}}
\newcommand{\zz}{\double{Z}}
\newcommand{\qqq}{\double{Q}}
\newcommand{\kk}{\double{K}}

\newcommand{\aaa}{\mathcal{A}}

\newcommand{\gggg}{\goth{g}}
\newcommand{\hhh}{\double{H}}
\newcommand{\mm}{\mathcal{M}}

\newcommand{\pp}{\pmatrix}

\newcommand{\dd}{\mathcal{D}}

\newcommand{\hh}{\mathcal{H}}

\newcommand{\ttt}{{\rm tr}}

\newcommand{\ot}{\otimes}
\newcommand{\op}{\oplus}

\newcommand{\bb}{\begin{eqnarray}}
\newcommand{\ee}{\end{eqnarray}}
\newcommand{\eee}{\nonumber\end{eqnarray}}
\newcommand{\qq}{\quad}
 
\begin{document}

\font\twelve=cmbx10 at 13pt
\font\eightrm=cmr8

\thispagestyle{empty}

\begin{center}

CENTRE DE PHYSIQUE TH\'EORIQUE $^1$ \\ CNRS--Luminy, Case
907\\ 13288 Marseille Cedex 9\\ FRANCE\\

\vspace{2cm}

{\Large\textbf{Krajewski diagrams and spin lifts}} \\

\vspace{1.5cm}

{\large Thomas Sch\"ucker $^2$}

\vspace{1.5cm}

{\large\textbf{Abstract}}
\end{center}

A classification of irreducible, dynamically non-degenerate, almost
commutative spectral triples is refined. It is extended to include
centrally extended spin lifts. Simultaneously it is reduced by imposing
three constraints: (i) the condition of vanishing Yang-Mills and mixed
gravitational anomalies, (ii) the condition that the fermion
representation be complex under the little group, while (iii) massless
fermions are to remain neutral under the little group. These
constraints single out the standard model with one generation of
leptons and quarks and with an arbitrary number of colours.

\vspace{1.5cm}

\vskip 1truecm

PACS-92: 11.15 Gauge field theories\\
\indent MSC-91: 81T13 Yang-Mills and other gauge theories

\vskip 1truecm

\noindent CPT-2005/P.003\\
\noindent hep-th/0501181

\vspace{2cm}
\noindent $^1$ Unit\'e Mixte de Recherche  (UMR 6207)
 du CNRS  et des Universit\'es Aix--Marseille 1 et 2 et  Sud
Toulon--Var, Laboratoire affili\'e \`a la FRUMAM (FR 2291)\\
$^2$ Also at Universit\'e Aix--Marseille 1, \\
 schucker@cpt.univ-mrs.fr 

\section{Introduction}

A small class of Yang-Mills theories with fermions admit an
interpretation in terms of almost commutative 4-dimensional
spectral triples \cite{book,real,grav,cc}. In these theories the
Yang-Mills forces are pseudo-forces induced by gravity
via transformations belonging to the automorphism group of the
algebra defining the noncommutative geometry. In the almost
commutative case this group consists of timespace
diffeomorphisms and gauge transformations. In the same way that
in general relativity the diffeomorphisms produce the
gravitational field, the gauge transformations produce a scalar
field, the metric of internal space. In the same vein, the
Einstein-Hilbert action can be generalized to noncommutative
geometry, the so-called spectral action. In the almost commutative
case it contains the Yang-Mills action. On the internal space the
spectral action reduces to the Higgs-potential and breaks the gauge
group spontaneously. The minimum of the Higgs-potential induces
the fermion masses and they
satisfy the Einstein equation in internal space. A natural question
to ask is whether these solutions are stable under  renormalization
flow. This question motivates our definition of dynamical
non-degeneracy \cite{class}. At this point we must note that the
standard model of electromagnetic, weak and strong forces
is a Yang-Mills theory that can be interpreted as a spectral
triple, that this interpretation produces the phenomenologically
correct Higgs field, a colorless isospin doublet, and that its
spectral triple is dynamically non-degenerate. Furthermore, after
restriction to one generation of leptons and quarks, this spectral
triple is irreducible. In
\cite{class} we started the classification of irreducible, dynamically
non-degenerate triples for finite algebras with one, two and three
simple summands. We simplified the task by lifting unitaries
instead of automorphisms. Lifting automorphism to the Hilbert
space is delicate for algebras $M_n(\cc)$, $n\ge 2$. Indeed all its
automorphisms connected to the identity are inner and form the
group $U(n)/U(1)$. When we want to lift such automorphisms to
the Hilbert space we encounter a continuous infinity of
multi-values parameterized by $U(1)$. We avoid this obstruction by
a central extension \cite{fare}. Redoing the classification
\cite{class} with centrally extended automorphisms rather than
unitaries implies two complications: (i) the extension is not
unique, (ii) there are more unitaries than extended
automorphisms, therefore a triple, which is dynamically
degenerate with unitaries might be non-degenerate with extended
automorphisms. To keep the classification manageable three
additional, physically motivated constraints are introduced. We only
keep those triples and central extensions, 
\begin{itemize}\item that are free of Yang-Mills anomalies and
free of mixed gravitational-Yang-Mills anomalies,
\item whose fermionic representation is complex under the little
group in each irreducible component,  
\item whose massless fermions are neutral under the little
group, i.e. transform trivially.
\end{itemize} 
Of course the standard model with an arbitrary number of colours 
satisfies these criteria and we will show that within noncommutative
geometry it is essentially unique as such.

\section{Statement of the result}

Consider a finite, real, $S^0$-real, irreducible
spectral triple whose algebra has one, two or three simple summands
and the extended lift as described below. Consider the list of all
Yang-Mills-Higgs models induced by these triples and lifts. Discard
all models that have (i) a dynamically
degenerate fermionic mass spectrum,
 (ii)  Yang-Mills or gravitational
anomalies, (iii) a fermion multiplet whose representation under
the little group is real or pseudo-real, or (iv) a neutrino transforming
non-trivially under the little group. The remaining models are the
following, $p$ is the number of colours, $p\ge 2$, the gauge group is on
the left-hand side of the arrow, the little group on the right-hand side:
\begin{description}
\item[$p=3,5...$]
\bb\frac{SU(2)\times U(1)\times SU(p)}{\zz_2\times \zz_p}&
\longrightarrow 
&\frac{U(1)\times SU(p)}{\zz_p}\eee
The left-handed fermions transform according to a multiplet 
${\bf 2}\ot {\bf p}$ with hypercharge $q/(2p)$ and a multiplet 
$ {\bf p}$ with hypercharge $-q/2$. The right-handed
fermions sit in two multiplets ${\bf p}$ with hypercharges
$q(1+p)/(2p)$ and $q(1-p)/(2p)$ and one singlet with hypercharge
$-q$, $q\in\qqq$. The elements in $\zz_2\times \zz_p$ are
embedded in the center of $ SU(2)\times U(1)\times SU(p)$ as
\bb \left( \exp{\frac{2\pi i k}{ 2  }}
\,1_2\,,\, \exp[2\pi i(pk-2\ell)/q]\,,\, \exp{\frac{2\pi i \ell}{ p 
}}\,1_p\right),\qq  k=0,1,\qq  \ell =0,1,...,p-1.\ee
The Higgs scalar transforms as an
$SU(2)$ doublet,
$SU(p)$ singlet and has hypercharge $-q/2$.

With the number of colours $p=3$, this is the standard model with one
generation of quarks and leptons. 

We also have in our list two submodels of the above model defined
by the subgroups
\bb\frac{SO(2)\times U(1)\times SU(p)}{\zz_2\times \zz_p}&
\longrightarrow 
&\frac{U(1)\times SU(p)}{\zz_p},\cr \cr \cr 
\frac{SU(2)\times U(1)\times SO(p)}{\zz_2}&
\longrightarrow 
&U(1)\times SO(p).\eee
They have the same particle content as the standard model, in the
first case only the $W^\pm$ bosons are missing, in the second case
roughly half the gluons are lost.
\item[$p=2,4...$]
\bb\frac{SU(2)\times U(1)\times SU(p)}{\zz_p}&
\longrightarrow 
&\frac{U(1)\times SU(p)}{\zz_p}\eee
with the same particle content as for odd $p$.  But now we have three
possible submodels:
\bb\frac{SO(2)\times U(1)\times SU(p)}{\zz_p}&
\longrightarrow 
&\frac{U(1)\times SU(p)}{\zz_p}\,,\cr \cr \cr 
{SU(2)\times U(1)\times SO(p)}&
\longrightarrow 
&{U(1)\times SO(p)},\cr \cr \cr 
\frac{SU(2)\times U(1)\times USp(p/2)}{\zz_2}&
\longrightarrow 
&\frac{U(1)\times USp(p/2)}{\zz_2}.\eee
\end{description}

\section{The set up}

Let ($\aaa,\hh,\dd, J,\varepsilon,\chi)$ be a real, $S^0$-real, finite
spectral triple. $\aaa$ is a finite dimensional real algebra
represented faithfully 
 on  a finite dimensional complex Hilbert space $\hh$ via $\rho$.
Four additional operators are defined on
$\hh$: the Dirac operator
$\dd$ is selfadjoint, the real structure (or charge conjugation) $J$ is
antiunitary, and the
$S^0$-real structure $\varepsilon$ and the chirality
$\chi$ are both unitary involutions. These operators  satisfy:
\bb
\hspace{-0.1cm}\bullet \hspace{2cm} J^2=1,\qq
[J,\dd]=[J,\chi]=[\varepsilon,\chi]=[\varepsilon,\dd]=0, \qq
\varepsilon J=-J
\varepsilon,\qq\dd\chi =-\chi \dd, \cr
[\chi,\rho(a)]=[\varepsilon,\rho(a)]=[\rho(a),J\rho(b)J^{-1}]=
[[\dd,\rho(a)],J\rho(b)J^{-1}]=0, \forall a,b \in \aaa.
\eee
$\bullet$ The chirality
 can be written as a finite sum $\chi
=\sum_i\rho(a_i)J\rho(b_i)J^{-1}.$  This condition is called
{\it orientability.}\\ 
$\bullet$ The intersection form
 $\cap_{ij}:=\ttt(\chi \,\rho (p_i) J \rho (p_j) J^{-1})$ is
non-degenerate,
$\rm{det}\,\cap\not=0$. The
$p_i$ are minimal rank projections in $\aaa$. This condition is
called {\it Poincar\'e duality}.\\
$\bullet$ The kernel of $\dd$ has no nontrivial $\aaa$-invariant
subspace. 

We choose a basis of $\hh$ such that the five operators take the
form
\bb 
\rho=
\pp{
\rho_L & 0 & 0 & 0 \cr  0 & \rho_R & 0 & 0 \cr   0 & 0 &
\overline{\rho_L^c}  & 0 \cr  0 & 0 & 0 & \overline{ \rho_R^c} },\qq
\dd=\pp{0&\mm&0&0\cr 
\mm^*&0&0&0\cr  0&0&0&\overline{\mm}\cr
0&0&\overline{\mm^*}&0},\eee
\bb J=\pp{0&0&1&0\cr 0&0&0&1\cr 1&0&0&0\cr  0&1&0&0}\circ
{\rm compl.\ conj.},\ 
\varepsilon =\pp{1&0&0&0\cr 0&1&0&0\cr 0&0&-1&0
\cr 0&0&0&-1},\ 
\chi =\pp{-1&0&0&0\cr 0&-1&0&0\cr 0&0&1&0\cr  0&0&0&1}.\eee
The algebra is a finite sum of simple algebras,
$\aaa =
\oplus_{i} M_{n_i}(\kk_i)$ and $\kk_i=\rr,\cc,\hhh$ where
$\hhh$ denotes the quaternions. Except for complex conjugation
in $M_n(\cc)$ and permutations of identical summands in the
algebra $\aaa$, every algebra automorphism
$\sigma
$  is inner, $\sigma (a)=uau^{-1}=:i_u(a)$ for a unitary $ u\in
U(\aaa)$. This unitary is ambiguous by any central unitary $u_c\in
U(\aaa)\cap{\rm Center}(\aaa)$, indeed $i_{u_cu}=i_u$.
$M_n(\rr)$ and $M_n(\hhh)$ do not have central unitaries close
to the identity. We therefore start with the complex case
$\aaa =
\oplus_{i} M_{n_i}(\cc)$. Since $M_1(\cc)=\cc$ has no
automorphisms close to the identity, we have to distinguish the
cases $n_i=1$ and $n_i\ge 2$, and write
\bb\aaa=\cc^M\oplus
\bigoplus_{k=1}^N M_{n_k}(\cc)\ \owns
a=(b_1,...b_M,c_1,...,c_N),\qq n_k\geq 2.
\label{algebra}\ee Its group of unitaries is
\bb U(\aaa)=U(1)^M\times
\matrix{N\cr \times\cr {k=1}} U(n_k)=:U(1)^M\times G_\aaa\
\owns\ u=(v_1,...,v_M,w_1,...,w_N),\ee
$G_\aaa$ is the group of `noncommutative unitaries'. The
subgroup of automorphisms of $\aaa$ connected to the identity is
the group of inner automorphisms,
Int$(\aaa)=G_\aaa/ZG_\aaa$ where $ZG_\aaa
=U(1)^N\owns (w_{c1}1_{n_1},...,w_{cN}1_{n_N})$ is the subgroup
of noncommutative, central unitaries, $w_{ck}$ is a root of the
determinant of $w_k$. The spin lift is a group homomorphism $L$
from the automorphism group of the algebra $\aaa$ (or its
connected component Int$(\aaa)$ for simplicity) into the group of
unitary operators on
$\hh$ satisfying
$[L(\sigma ),J]=[L(\sigma ),\chi ]=0$ for all $\sigma \in {\rm
Int}(\aaa)$  and satisfying the covariance condition
\bb i_{L(\sigma )} \rho (a)= \rho (\sigma (a))\qq {\rm for \ all}
\qq a\in \aaa.\ee
 The ambiguity of the lift by the noncommutative, central unitaries
forces us to centrally extend Int$(\aaa)$ to $G_\aaa$. Then the
group homomorphism
$L:G_\aaa\rightarrow U(\hh)$ must satisfy
$[L(w ),J]=[L(w ),\chi ]=0$ for all $w \in G_\aaa$  and 
\bb i_{L(w )} \rho (a)= \rho (i_w (a))\qq {\rm for \ all} \qq a\in
\aaa.\ee
 The following map qualifies as extended lift \cite{fare}: 
\bb L(w)&=&\rho (\hat u)J\rho (\hat u)J^{-1},\cr \cr 
&&\hat v_j:=\prod_{k=1}^N(\det w_{k})^{ q_{j,k}},\qq
j=1,...,M,\cr 
&&\hat w_\ell:=w_\ell \prod_{k=1}^N(\det w_{k})^{
q_{{M+\ell},k}},\qq \ell=1,...,N,\label{generallift}\ee
 where the
$q$'s are arbitrary rational numbers, `charges'. They form a
$(M+N)\times N$ matrix. This lift is multi-valued, but thanks to the
central extension it only has a finite number of values. The maps $L$
are not the only possible extensions, if the representation decomposes
into several blocks different charges may be chosen in each block. 

The induced Yang-Mills model has $G_\aaa$ as gauge group and
the (extended) lift $L$ defines the fermionic representation. We will use
the conditions of vanishing Yang-Mills and mixed
gravitational-Yang-Mills anomalies
to reduce the possible charges in the definition of the lift. To spell
out these conditions for the general $L(w)$, equation
(\ref{generallift}), we need its infinitesimal version, $	\ell (X)$
defined by 
\bb L(\exp X)=\exp\ell (X),\qq X= (X_1,...,X_N) \in
\bigoplus ^N_ {k=1} u(n_k)=:\gggg_\aaa.\ee 
The vanishing of the Yang-Mills anomalies and of the
gravitational-Yang-Mills anomalies is equivalent to
\bb \ttt[ \ell (X)^3\chi (1+\varepsilon)/2 ]=0 \qq {\rm and}\qq
\ttt[ \ell (X)\chi (1+\varepsilon)/2 ]=0,\qq {\rm for\ all}\ X\in
\gggg_\aaa.\ee
Let us write the lift in components:
\bb L(w)&=:&\pp{L_L(w)&0&0&0\cr 0&L_R(w)&0&0\cr 
0&0&\bar L_L(w)&0\cr 0&0&0&\bar L_R(w)}\qq{\rm and}\\
\ell(X)&=:&\pp{\ell_L(X)&0&0&0\cr 0&\ell_R(X)&0&0\cr 
0&0&\bar \ell_L(X)&0\cr 0&0&0&\bar \ell_R(X)}.\ee
 Then the
anomaly conditions read:
\bb \ttt[ -\ell _L(X)^3+ \ell _R(X)^3]=0, \qq 
\ttt[ -\ell _L(X)+ \ell _R(X)]=0,\qq {\rm for\ all}\ X\in
\gggg_\aaa.\ee

 The scalar field is obtained by fluctuating
the internal Dirac operator $\dd$, that is a finite linear
combination:
\bb
\Phi  :=\sum_j r_j\,L(_jw ) \, \dd \, L(_jw)^{-1},\qq
r_j\in\rr,\
_jw \in G_\aaa.\ee
After the decomposition
\bb\Phi =:\pp{0&\phi &0&0\cr \phi ^*&0&0&0\cr 
0&0&0&\bar\phi \cr 0&0&\bar\phi ^*&0},\ee
we have
\bb \phi  :=\sum_j r_j\,L_L(_jw ) \, \mm \, L_R(_jw)^{-1}.\ee

The action of the scalar field is the Higgs potential
\bb V(\Phi  )= \lambda\  \ttt\left[ \Phi ^4\right]
-\textstyle{\frac{\mu ^2}{2}}\
\ttt\left[
\Phi  ^2\right]
= 4\lambda\  \ttt\left[ (\phi^*\phi ) ^2\right]
-2{\mu ^2}\
\ttt\left[
\phi^*\phi\right] ,\ee
 where $\lambda $ and $\mu $ are positive
constants
\cite{cc,kraj2}.
Our task is to find the minima $\stackrel{\circ}{\Phi } $ of this
action, their spectra and their  {\it little groups}
\bb G_\ell:=\left\{ w \in G_\aaa,\ L(w) 
\stackrel{\circ}{\Phi }  L(w)^{-1}=\ \stackrel{\circ}{\Phi }
\right\} .\ee

If we replace one of the $M_{n_k}(\cc)$, $n_k\ge 2$, by
$M_{n_k}(\rr)$ or
$M_{n_k}(\hhh)$ in the algebra (\ref{algebra}) then
the lift (\ref{generallift}) simplifies: 
$q_{M+k,j}=0$ for all
$1\le j\le N$ and $q_{j,k}=0$ for all  $1\le j\le M+ N$. If we replace
say the $\kk_1=\cc$ by $\kk_1=\rr$ then $q_{1,j}=0$ for all
$1\le j\le  N$.

In \cite{class} we had lifted the entire unitary group
$U(\aaa)\owns u$ by means of the lift $L(u)=\rho (u)J\rho
(u)J^{-1}$ which up to phases coincides with the present lift
(\ref{generallift}). In all cases where $\aaa$ has no commutative
summand, $\cc$ or $\rr$, which means $M=0$, the phases can be
re-absorbed into the noncommutative unitaries and there is no
modification to the configuration space, the affine space of all
scalars. In the following we go through all irreducible Krajewski
diagrams with
$N+M\le 3$ as given in  \cite{class} using the most general
{\it anomaly free} lift (\ref{generallift}) and work out the
modifications for $M\ge 1$.  All
induced Yang-Mills-Higgs models have one massless particle, a
`neutrino'. Our aim is to list all those models whose fermion masses are
dynamically non-degenerate, whose fermions transform as complex
representations under the little group in each irreducible component,
and whose neutrino is neutral under the little group.

Let us briefly recall the definition of dynamical degeneracy
\cite{class}. The mass matrix $\mm$ decomposes into blocks, which are
represented by the arrows in the Krajewski diagram. Each arrow comes
with an orientation and three algebras, a left-handed algebra, a
right-handed algebra and a colour algebra. If the sizes of the
matrices, elements of the three algebras, are $k$, $\ell$ and $p$, then
the block corresponding to this arrow is $M_\cdot\ot 1_p$ with
$M_\cdot $ being a complex $k\times\ell$ matrix. The spectrum of the
(internal) Dirac operator $\dd$ is
always degenerate: all nonvanishing eigenvalues come in pairs of
opposite sign due to the chirality that anticommutes with $\dd$, {\it
`left-right degeneracy'}. All eigenvalues appear twice due to the real
structure that commutes with
$\dd$, {\it `particle anti-particle degeneracy'}. There is a third
degeneracy, $p$-fold for the block above, that comes from the first
order axiom. Let us call it {\it colour degeneracy}. It is absent if and
only if the colour algebras of all arrows are commutative. We call these
degeneracies {\it kinematical} because these come from the axioms.
Because of the axioms, these three degeneracies survive the
fluctuations of the Dirac operator and the minimization of the Higgs
potential. By dynamical non-degenerate we mean that no minimum of
the Higgs potential has degeneracies other than the above three. The
first two degeneracies survive {\it quantum fluctuations} as well. We
also want the colour degeneracies to be protected from quantum
fluctuations. A natural protection is unbroken gauge invariance, a
requirement that we include in the definition of dynamical
non-degeneracy.  More precisely, the
irreducible
 spectral triple $(\aaa,\hh,\dd, J, \epsilon ,\chi )$ is  dynamically
non-degenerate if all minima $\stackrel{\circ}{\Phi }$ of the Higgs
potential
 define  again a spectral triple $(\aaa,\hh,\stackrel{\circ}{\Phi }, J,
\epsilon ,\chi )$ and if the spectra of all minima  have no degeneracies
other than the three kinematical ones. We also suppose that the colour
degeneracies are protected by the little group. By this we mean that all
eigenvectors of
$\stackrel{\circ}{\Phi }$ corresponding to the same eigenvalue are in
a common orbit of the little group (and scalar multiplication and
charge conjugation).

Recall also that a unitary representation is real if it is equal to its
complex conjugate and pseudo-real if it is unitarily equivalent to its
complex conjugate. Otherwise the representation is complex. For
example the representations of $SO(n)$ are real, those of
$SU(2)$ are pseudo-real. An irreducible, unitary representation of
$U(1)$ is complex if and only if its charge is non-zero. The physical
motivation for complex representations is that they allow to
distinguish between particles and anti-particles.

\section{One and two summands}

Of all simple algebras only $\aaa=M_n(\cc)$ admit real spectral
triples \cite{kps}. The irreducible triples with $n\ge 2$ are
dynamically degenerate. If
$n=1$ then $\aaa=\cc$. Being commutative this algebra has no
automorphisms close to the identity and there is nothing to extend
nor to lift.

For the sum of two algebras only $\aaa = M_2(\cc)\op \cc\owns
(a,b)$ and its real or quaternionic subalgebras admit  dynamically
non-degenerate triples. These are of the form
\bb \rho (a,b)=\pp{a&0&0&0\cr 0&^\beta\bar b &0&0\cr 
0&0& b&0\cr 0&0&0& b},\qq\beta =\pm 1,\ ^1 b:= b,\ 
^{-1} b:=\bar b,\qq
\mm=\pp{0\cr m},\qq m\in\rr.\ee
We have one noncommutative unitary $w\in U(2)$ and the lift
(\ref{generallift}) reduces to
\bb L(w)=\rho (w (\det w)^{q_{21}},(\det w)^{q_{11}})J\rho (w (\det
w)^{q_{21}},(\det w)^{q_{11}})J^{-1}\ee
or equivalently 
\bb  L_L(w)=w(\det w)^{q_{21}-q_{11}},\ 
L_R(w)=w(\det w)^{-(\beta +1)q_{11}}.\ee
On infinitesimal level with $X\in u(3)$ and
$X_0:=X-{\textstyle\frac{1}{2}} \ttt X\,1_2$ we have
\bb\ell_L=X_0+(q_{21}-q_{11}+{\textstyle\frac{1}{2}} )\,\ttt X\,1_2,\qq
\ell_R=-(\beta +1)p\,\ttt X\,1_2\,,\ee
and the lift is anomaly free if and only if $q_{21}=-1/2$, $q_{11}=0$ for
the representation $\beta =1$ and $q_{21}=q_{11}-1/2$ for $\beta =-1$.
Therefore all fermionic hypercharges vanish and the gauge group is
$SU(2)$. The scalar field is a doublet, $\phi =(x,y)^T,\
x,y\in\cc$ and $\stackrel{\circ}{\phi
}\ =(\mu/(4\lambda )^{1/2},0)^T$ minimizes the Higgs potential
with little group
$G_\ell=\{1_2\}$. 
Replacing $M_2(\cc)$ by the quaternions $\hhh$ leads to the same
Yang-Mill-Higgs model: $SU(2)\rightarrow \{1_2\}$ with a
left-handed doublet of fermions, a right-handed singlet and a
doublet of scalars. If we take $M_2(\rr)$ then only the gauge group
changes: $SO(2)\rightarrow \{1_2\}$. In all cases the little group
is trivial and has no complex representation.

\section{Three summands}

We use the list of the 41 irreducible Krajewski diagrams with three
simple summands from
\cite{class}, figure 1. 
This list becomes exhaustive upon permutations of the three algebras
$\aaa_1=M_n(\kk_1)\owns a$, $\aaa_2=M_m(\kk_2)\owns b$,
$\aaa_3=M_q(\kk_3)\owns c$, upon permuting left and right, i.e.
changing the directions of all arrows simultaneously, and
upon permutations of particles and antiparticles independently in
every connected component of the diagram.

Let  $k$, $\ell$, $p$ be the sizes of the matrices $a,$ $b,$ $c$, for example
$k=n$ if $\kk_1=\cc$ or $\rr$, $k=2n$ if $\kk_1=\hhh$. To write the lift
we will use the following letters:
\bb \hat u&:=&u\,(\det u)^{q_{11}}(\det v)^{q_{12}}(\det
w)^{q_{13}}\in U(\aaa_1),\\
\hat v&:=&v\,(\det u)^{q_{21}}(\det v)^{q_{22}}(\det w)^{q_{23}}\in
U(\aaa_2),\\
\hat w&:=&w\,(\det u)^{q_{31}}(\det v)^{q_{32}}(\det w)^{q_{33}}
\in U(\aaa_3) .\ee
It is understood that for instance if $k=1$ we set $u=1$ and $q_{j1}=0$, 
$j=1,2,3$. If
$\kk_1=\hhh$ or $\rr$ we set $q_{j1}=0$ and
$q_{1j}=0$. 
\vspace{1\baselineskip}

{\bf Diagram 1} yields:
\bb \rho _L (a,b,c)=\pp{a\ot 1_k&0\cr  0&b\ot 1_\ell},&&
\rho _R(a,b,c)=\pp{^\beta b\ot 1_k&0\cr  0&c\ot 1_\ell},\cr \cr \cr
\rho _L^c(a,b,c)=\pp{1_k\ot\,^{\alpha '} a&0\cr  0&1_\ell\ot\,^{\beta '}
b},&&
\rho _R^c(a,b,c)=\pp{1_\ell\ot\,^{\alpha '} a&0\cr  0&1_p\ot\,^{\beta '}
b},\ee
where $\beta ,\alpha ',\beta '$ take values $\pm 1$ to
indicate whether the fundamental representation, 1, or its
complex conjugate, $-1$, is meant. The fermionic mass matrix has the
form
\bb \mm=\pp{M_1\ot 1_k&0\cr  1_\ell\ot M_3&M_2\ot1_\ell},\qq M_1\in
M_{k\times
\ell}(\cc),\ M_2\in M_{\ell\times p}(\cc),\ M_3\in M_{\ell\times
k}(\cc).\ee 
If $M_3$ is nonzero the first order axiom implies $\beta =1$. We consider
only this case, the other case, $M_3=0$ is treated as diagram 2. 
 The first two summands, $\aaa_1$ and
$\aaa_2$ are colour algebras, they are both broken and must therefore
be 1-dimensional,
$k=\ell=1$. Then we must take $p=2$, for $p\ge 3$ we would have two or
more neutrinos.

Vanishing anomalies imply $q_{33}+1/2+\beta 'q_{23}=0,$
$q_{13}=-\beta 'q_{23}$. The little group then is $U(1)$ or trivial, but in
the former case the neutrino is charged.

\vspace{1\baselineskip}

{\bf Diagram 2} yields:
\bb \rho _L (a,b,c)=\pp{a\ot 1_k&0\cr  0&c\ot 1_\ell},&&
\rho _R(a,b,c)=\pp{b\ot 1_k&0\cr  0&b\ot 1_\ell},\cr \cr \cr
\rho _L^c(a,b,c)=\pp{1_k\ot a&0\cr  0&1_p\ot b},&&
\rho _R^c(a,b,c)=\pp{1_\ell\ot a&0\cr  0&1_\ell\ot b},\ee
 and
\bb \mm=\pp{M_1\ot 1_k&0\cr  0&M_2\ot1_\ell},\qq M_1\in M_{k\times
\ell}(\cc),\ M_2\in M_{p\times \ell}(\cc).\ee 
The possible complex conjugations in the representations are irrelevant
in this diagram. As in diagram 1 be must take $k=\ell=1$, $p=2$.

Let us write the fluctuations in the form:
\bb \phi =\pp{\varphi _1\ot 1_k&0\cr  0&\varphi _2\ot1_\ell},\qq
\varphi _1\in M_{k\times
\ell}(\cc),\ \varphi _2\in M_{p\times \ell}(\cc)\ee
with 
\bb\varphi _1  =\sum_j r_j\,\hat u _jM_1 \hat v_j^{-1},\qq
\varphi _2  =\sum_j r_j\,\hat w _jM_2 \hat v_j^{-1},\ee

We can decouple the two scalars $\varphi _1$ and $\varphi
_2$ by means of the fluctuation:
$ r_1={\textstyle\frac{1}{2}},$ $\hat u_1=1_k,$ $\hat v_1=1_\ell,$
$\hat w_1=1_p$,
$ r_2={\textstyle\frac{1}{2}},$ $\hat u_2=1_k,$ $\hat v_2=1_\ell,$
$\hat w_2=-1_p$. Note that $\hat w_2=-1_2$ is possible because 
det$(-1_2)=1$.
Since the arrows $M_1$ and $M_2$ are disconnected, the Higgs potential
is a sum of a potential in $\varphi _1$ and of a potential in $\varphi _2$.
Proceeding as in the preceding section we find the minimum
$\stackrel{\circ}{\varphi }_1=\mu{(4\lambda})^{-\frac{1}{2}}$  and
$\stackrel{\circ}{\varphi }_2$ has one eigenvalue
$\mu{(4\lambda})^{-\frac{1}{2}}$ and one vanishing eigenvalue. All
triples associated to the first diagram are therefore dynamically
degenerate.

Similarly, we can discard {\bf diagrams 3, 4, 6. } 
\vspace{1\baselineskip}

{\bf Diagram 5} yields:
\bb \rho _L (a,b,c)=\pp{a\ot 1_k&0\cr  0&b\ot 1_p},&&
\rho _R(a,b,c)=b\ot 1_k,\cr \cr \cr
\rho _L^c(a,b,c)=\pp{1_k\ot a&0\cr  0&1_\ell\ot c},&&
\rho _R^c(a,b,c)=1_\ell\ot a,\eee \vspace{-0.5cm}
\bb \mm=\pp{M_1\ot 1_k\cr  1_\ell\ot M_2},\qq M_1\in M_{k\times
\ell}(\cc),\ M_2\in M_{p\times k}(\cc).\ee
The colour algebra is indexed by $k$ and $\ell$. Both summands are
broken, therefore $k=\ell=1$ forcing $p=1$ to avoid two or more neutrinos.
\vspace{1\baselineskip}

{\bf Diagrams 7, 9, 11, 12} fall in the same way.
\vspace{1\baselineskip}

{\bf Diagram 8} yields the representations
\bb \rho _L(a,b,c) =\pp{ a\ot 1_k&0&0\cr  0& c\ot
1_k&0\cr  0& 0&b\ot 1_p},\ \rho _R(a,b,c)=\pp{^\beta b\ot 1_k&0\cr 
0&^{\gamma }c\ot 1_p},\ee
\bb\rho _L^c(a,b,c)=\pp{1_k\ot\,^{\alpha '} a&0&0\cr 
0&1_p\ot\,^{\alpha '} a&0\cr  0& 0&1_\ell\ot \,^{\gamma '}c},\ 
\rho _R^c(a,b,c)=\pp{1_\ell\ot\,^{\alpha '} a&0\cr 
0&1_p\ot\,^{\gamma '} c},\ee
where $\beta ,\alpha ',\gamma '$ take values $\pm 1$ to
indicate whether the fundamental representation, 1, or its
complex conjugate, $-1$, is meant. The primes refer to colour
representations. These leave the Higgs scalars invariant.  The mass
matrix is
\bb \mm=\pp{M_1\ot 1_k&0\cr  M_2\ot 1_k&0\cr  0&{M_3}^*\ot
1_p},\qq M_1\in M_{k\times\ell}(\cc),\qq  M_2,M_3\in
M_{p\times\ell}(\cc).\ee 
Requiring at most one zero eigenvalue (up to a
possible colour degeneracy)  implies $k=1$, $\ell=p+1$ or $k=1$,
$\ell=p$. The colour group consists of the
$u$s and $w$s. As they are spontaneously broken we must have
$k=p=1$, leaving $\ell=2$. Then the  fluctuations are
\bb
\varphi _1  =\sum_j r_j\,\hat u _jM_1 \,^\beta \hat v_j^{-1},\qq
\varphi _2  =\sum_j r_j\,\hat w _jM_2 \,^\beta \hat v_j^{-1},\qq 
\varphi _3  =\sum_j r_j\,\hat w _jM_3 \hat v_j^{-1},\ee
 where 
\bb \hat u= (\det v)^{q_{12}},\qq
\hat v= v\,(\det v)^{q_{22}},\qq
\hat w= (\det v)^{q_{32}}.\ee 
With
$C_i:={\varphi _i}^* \, \varphi _i$ the Higgs potential reads
\bb V(C_1,C_2,C_3)=4  [\lambda\, {\rm tr} (C_1+C_2)^2  -
{\textstyle\frac{1}{2}} \mu ^2\, {\rm tr} (C_1+C_2)] + 4  [\lambda\,
{\rm tr} (C_3)^2- {\textstyle\frac{1}{2}} \mu ^2 \,{\rm tr} (C_3)].
\ee  
If we impose an anomaly free lift then in all but four cases the little
group is trivial,
$G_\ell=\{1_2\}$. The exceptions have $\kk_1=\kk_2=\kk_3=\cc$, i.e.
$\aaa=\cc\op M_2(\cc)\op\cc$, $q_{12}=-q_{32},\ q_{22}=-1/2$ and
$\beta ,\gamma ,\gamma ',\alpha '=---+$, $+--+$, $-++-$ or $+++-$.
These four triples induce the electro-weak model
$(SU(2)\times U(1))/\zz_2\rightarrow U(1)$ of protons, neutrons,
neutrinos and electrons. One chiral part of the neutron is an $SU(2)$
singlet and therefore a real representation under the little group.  It is
nevertheless amusing to note a mass relation in these models: in the first
and the third triple  the neutron is slightly heavier than the proton,
\bb m_p=m_n\,\sqrt{\left( 1+\,\frac{m_e^4}{m_n^4}\,\right)/  
\left( 1+\,\frac{m_e^2}{m_n^2}\,\right)},\label{pn}\ee
in the second and fourth triple
the neutron is slightly lighter than the proton,
\bb m_n=m_p\,\sqrt{\left( 1+\,\frac{m_e^4}{m_p^4}\,\right)/  
\left( 1+\,\frac{m_e^2}{m_p^2}\,\right)}.\label{np}\ee
\vspace{1\baselineskip}

{\bf Diagram 10} is similar to diagram 8. It needs $k=p=1$ and
$\ell=2$ and has representations
\bb \rho _L (a,b,c) =\pp{a&0&0\cr  0&^\alpha a&0\cr  0& 0&b
},&&
\rho _R(a,b,c) =\pp{^\beta b&0\cr  0&c},\\ \cr \cr 
\rho _L^c(a,b,c) =\pp{^{\alpha '} a&0&0\cr  0&^{\alpha '} a&0\cr  0&
0&^{\gamma '} c1_2},&&
\rho _R^c(a,b,c) =\pp{^{\alpha '} a1_2&0\cr  0&^{\gamma '}
c}.\ee
 The mass matrix and Higgs potential are as for diagram 8, while the
  fluctuations read
\bb
\varphi _1  =\sum_j r_j\,\hat u _jM_1\,^\beta  \hat v_j^{-1},\qq
\varphi _2  =\sum_j r_j\,^\alpha \hat u _jM_2\,^\beta  \hat v_j^{-1},\qq 
\varphi _3  =\sum_j r_j\,\hat w _jM_3 \hat v_j^{-1}.\ee
Only four triples are anomaly free and have non-trivial little group:
they all have
$\kk_1=\kk_2=\kk_3=\cc$ and $q_{22}=-1/2$ and are given by
(i) 
$\alpha ,\beta ,\alpha ' ,\gamma '=---+$, $q_{12}=-q_{32}$,
(ii) $--++$, $q_{12}=q_{32}$,
(iii) $-+-+$, $q_{12}=-q_{32}$,
(iv) $-+++$, $q_{12}=q_{32}$.
 As before they induce the electro-weak model with one mass relation.
\vspace{1\baselineskip}

{\bf Diagram 13} has {\it ladder form}, i.e. it consists of horizontal
arrows,  vertically aligned.  Its representations are
\bb \rho _L (a,b,c) =\pp{a\ot 1_k&0&0\cr  0&^{\alpha _1}a\ot 1_k&0\cr 
0& 0&^{\alpha _2}a\ot 1_p},&&
\rho _R (a,b,c) =\pp{b\ot 1_k&0\cr  0&^\beta b\ot 1_p},\\ \cr \cr
\rho _L^c (a,b,c) =\pp{1_k\ot\,^{\alpha '} a&0&0\cr  0&1_k\ot\,^{\alpha
'} a&0\cr  0& 0&1_k\ot c},&&
\rho _R^c (a,b,c) =\pp{1_\ell\ot\,^{\alpha '} a&0\cr  0&1_\ell\ot c}.\ee
The mass matrix is
\bb \mm=\pp{M_1\ot 1_k&0\cr  M_2\ot 1_k&0\cr  0&M_3\ot 1_p},\qq
M_1, M_2,M_3\in M_{k\times\ell}(\cc).\ee
 The fluctuations read
\bb
\varphi _1  =\sum_j r_j\,\hat u _jM_1\,^\beta  \hat v_j^{-1},\qq
\varphi _2  =\sum_j r_j\,^{\alpha_1} \hat u _jM_2  \hat
v_j^{-1},\qq 
\varphi _3  =\sum_j r_j\,^{\alpha 2} \hat u _jM_3\,^\beta  \hat
v_j^{-1},\ee
 and the action is
\bb V(C_1,C_2,C_3) = 4k \, [\lambda\, {\rm tr} (C_1+C_2)^2 - 
{\textstyle\frac{1}{2}} \mu ^2\, {\rm tr} (C_1+C_2)] + 4p \, [\lambda\,
{\rm tr} (C_3)^2 - {\textstyle\frac{1}{2}} \mu ^2 \,{\rm tr} (C_3)].
\ee
 The neutrino count  implies $k=1$, $\ell=1$ or 2.

{\bf 1:} If $\ell =1$ we must take $\kk_1=\cc$ and 
$\alpha _1=-1$, otherwise the kernel of $\dd$ would have a nontrivial
$\aaa$-invariant subspace. We also must take $\kk_3=\cc$ and $p\ge 2$,
otherwise there would be no automorphism to lift nor to extend. Then
the extended lift reduces to
\bb \hat u= (\det w)^{q_{13}},\qq
\hat v= (\det w)^{q_{23}},\qq
\hat w=w\, (\det w)^{q_{33}}.\ee 
 All remaining triples that admit an
anomaly free extension have little group $SU(p)$. Its representation on
the first fermion is trivial.

{\bf 2:} If $\ell =2$ we must take $\kk_1=\rr$ or $\cc$ and  $p=1$,
otherwise the neutrino would have unbroken colour. The lift has
\bb \hat u= (\det v)^{q_{12}},\qq
\hat v= v\,(\det v)^{q_{22}},\qq
\hat w= (\det v)^{q_{32}}.\ee 
All anomaly free lifts have a trivial little group.
\vspace{1\baselineskip}

{\bf Diagram 18} goes down the same drain.
{\bf Diagrams 14, 16, 19, 21} must have $\ell=1$ and $k=1$ or 2. The first
case falls as diagram 13 with $\ell=1$, the second has broken colour.
{\bf Diagrams 15, 20, 23 and 24} must have $k=\ell=1$ and are rejected as
diagram 13 with $\ell=1$.
\vspace{1\baselineskip}

{\bf Diagram 17} provides  models satisfying all criteria, in
particular the standard model. In order to obtain the latter in
conventional physics notations, `left things left', we interchange left
with right and $\aaa_1$ with $\aaa_2$. Then with $(a,b,c)\in
M_k(\cc)\op M_\ell(\cc)\op M_p(\cc)$ the representation reads:
\bb
\rho _L=\pp{a\ot 1_p &0\cr 0&^\alpha a\ot 1_\ell},&&
\rho _R=\pp{b\ot 1_p&0&0\cr 0&^{\beta _1}b\ot 1_p&0\cr 
0&0&^{\beta _2}b\ot 1_\ell},\\\cr \cr 
\rho _L^c=\pp{1_k\ot c&0\cr 0&1_k\ot\,^{\beta '}b},&&
\rho _R^c=\pp{1_\ell\ot c&0&0\cr 0&1_\ell\ot c&0\cr 
0&0&1_\ell\ot\, ^{\beta '}b}.\ee
The mass matrix is:
\bb\mm=\pp{M_1\ot 1_p&M_2\ot 1_p&0\cr 0&0&M_3\ot 1_\ell}.\ee
To avoid more than one neutrino we must take $k=1$ or 2 and $\ell=1$.
The first case forces $p\ge 2$ implying that the neutrino has unbroken
colour. Therefore we take $k=2$ and $\ell=1$. If $p\ge 2$ we have
\bb \hat u = u \det u^{q_{11}}\,\det w^{q_{13}},\qq
\hat v =\det u^{q_{21}}\,\det w^{q_{23}},\qq
\hat w = w \det u^{q_{31}}\,\det w^{q_{33}}.\ee
Only four sign choices, $\alpha ,\beta _1,\beta _2,\beta '=+---,\ +-++,
\ --++$ and $----$, admit an anomaly free lift with complex fermion
representations under the little group. All four choices lead to the same
model, 
\bb \,\frac{SU(2)\times U(1)\times SU(p)}{\zz_2\times \zz_p}&
\longrightarrow 
&\frac{U(1)\times SU(p)}{\zz_p}\, ,\qq p=3,5,...,\cr \cr \cr 
\,\frac{SU(2)\times U(1)\times SU(p)}{\zz_p}&
\longrightarrow 
&\frac{U(1)\times SU(p)}{\zz_p}\, ,\qq p=2,4,... \eee
For example with the first choice we get:
\bb q_{11}=-\,\frac{1}{2}\,  ,& q_{21}=\,\frac{x}{2}\, ,&
q_{31}=\,\frac{x}{2p}\, ,\\
q_{13}=\,0\,  ,\qq& q_{23}=\,\frac{y}{2}\, ,&
q_{33}=\,\frac{y}{2p}\, -\,\frac{1}{p}\, ,\ee
where $x$ and $y$ are rational numbers, $x+y\not= 0$.
The hypercharges of the five irreducible fermion
representations  under the gauge group are:
\bb \,\frac{x+y}{2p}\,,\qq -\,\frac{x+y}{2}\,;\qq\qq
(x+y)\,\frac{1+p}{2p}\,,\qq (x+y)\,\frac{1-p}{2p}\,,\qq
-(x+y).\ee 

 A few other choices of the fields $\kk_1,\kk_2,\kk_3$ induce
submodels of the above one that still satisfy all of our criteria:\\
$\hhh,\cc,\cc$ entails $\alpha =1$, $q_{i1}=0,\ i=1,2,3$, $x=1$ and
reproduces the standard model, $SU(2)\times U(1)\times SU(p)
\longrightarrow {U(1)\times SU(p)},$ where to alleviate notations we
suppress the quotient by discrete groups.\\
$\rr, \cc,\cc$ entails $\alpha =1$, $q_{i1}=0,\ i=1,2,3$, $x=1$ and
induces the standard model with the $W$s missing, ${U(1)\times
U(1)\times SU(p)}\longrightarrow {U(1)\times SU(p)}.$\\
$\cc, \cc,\hhh$ entails $q_{i3}=0,\ i=1,2,3$, $y=1$ and
induces the standard model with a few gluons, roughly half of them,
missing,
${SU(2)\times U(1)\times USp(p/2)}\longrightarrow
{U(1)\times USp(p/2)},$\\ $p$ even.\\
$\cc, \cc,\rr$ entails $q_{i3}=0,\ i=1,2,3$, $y=1$ and
induces the standard model with  some, roughly half, the gluons
missing,
${SU(2)\times U(1)\times SO(p)}\longrightarrow 
{U(1)\times SO(p)}.$\\
All other possibilities have all fermion hypercharges vanish.

 Finally for $p=1$ we have $\kk_i=\cc,\ i=1,2,3$ and $q_{13}=0,\ 
q_{23}=0,\  q_{33}=-1$. The only models satisfying our criteria have
again  $\alpha ,\beta _1,\beta _2,\beta '=+---,\ +-++,
\ --++$ or $----$. These four triples induce the electro-weak model
$SU(2)\times U(1)\rightarrow U(1)$ of protons, neutrons, neutrinos and
electrons, this time without a mass relation. E.g. for the first
choice, we have $q_{11}=-1/2,\  q_{21}=x/2,\ q_{31}=x/2.$

{\bf Diagram 22} induces the same models as diagram 17.

The following diagrams can be excluded simply by imposing broken
colour to be 1-dimensional and by requiring the model to have at most
one neutrino: {\bf diagrams 25, 26, 27, 28, 31, 32, 33.} 

{\bf Diagram 29} is similar to diagram 8. It takes $k=p=1$ and
$\ell=2$. The representations are
\bb \rho _L =\pp{b&0\cr  0&^\alpha a
},&&
\rho _R =\pp{\bar a&0&0\cr  0&c&0\cr 0&0&^\beta b},\\ \cr  
\rho _L^c =\pp{^{\alpha } a1_2&0\cr  0&\bar c},&&
\rho _R^c =\pp{^{\alpha } a&0&0\cr  0&^\alpha a&0\cr 0&0&\bar c}\ee
 with a mass matrix
\bb \mm=\pp{M_1&M_2&0\cr 0&0&M_3^*}.\ee
Only three triples are anomaly free and have non-trivial little group:
they all have
$\kk_1=\kk_2=\kk_3=\cc$ and are given by
(i) 
$\alpha ,\beta =--$, $q_{12}=q_{32}$, $q_{22}=-1/2$,
(ii) $-+$, $q_{12}=q_{32}$, $q_{22}=-1/2$,
(iii) $+-$, $q_{22}=-1/2-q_{12}/2$, $q_{32}=0$. 
 As before they induce the electro-weak model with one mass relation.
\vspace{1\baselineskip}

{\bf Diagram 30} takes $k=\ell=1$ and
$p=2$. The representations are
\bb \rho _L =\pp{c&0\cr  0&^\alpha a
},&&
\rho _R =\pp{\bar b&0&0\cr  0&\bar b&0\cr 0&0&^\gamma c},\\ \cr  
\rho _L^c =\pp{ a1_2&0\cr  0&^\alpha b},&&
\rho _R^c =\pp{ a&0&0\cr  0&a&0\cr 0&0&^\alpha b1_2}\ee
 with a mass matrix
\bb \mm=\pp{M_1&M_2&0\cr 0&0&M_3^*}.\ee
Note that in this model $M_1$ and $M_2$ fluctuate in the same way.
This entails that the little group is trivial 
\vspace{1\baselineskip}

For {\bf diagram 34}, broken colour and neutrino count  imply
$k=\ell=1$. All remaining triples with extended, but anomaly free lift
have a non-trivial invariant subspace in the kernel of the Dirac
operator or a real representation under the little group. 

{\bf Diagrams 35, 38 and 39} share this fate.

{\bf Diagram 36, 37, 40 and 41} must have $k=\ell=1$ for broken colour
and neutrino count and $p=1$ to avoid a coloured neutrino.

At this point we have exhausted all irreducible triples for algebras with
one, two and three simple summands.

\section{Outlook}

We conjectured \cite{class} that for five and more summands there is
no irreducible, dynamically  non-degenerate spectral triple without
mass relations and that for four summands we only have the standard
model with two simple colour algebras. Jureit and Stephan have written
a computer algorithm that allows to compute the irreducible Krajewski
diagrams with four summands and letter changing arrows. So far the
conjecture for four summands resists their findings.  The extension of
the above list of Yang-Mills-Higgs models to four algebras is also in
work. 

The aim is of course to realize the old dream of Grand Unification
\cite{gg} within noncommutative geometry. The dream was to describe
particle physics by means of a Yang-Mills-Higgs model with (i) a
simple group or at least a group as simple as possible, (ii) an irreducible
fermion representation, or at least irreducible in one generation of
quarks and leptons; that the fermion representation be (iii) anomaly
free and (iv) complex. Let us mention two main differences between the
two approaches. First,  being an extension of
Riemannian geometry, noncommutative geometry unifies the standard
model of electromagnetic, weak and strong forces with gravity already
at non-quantum level. Second, while Grand Unification predicts new
forces at the unification scale of $10^{17}$ GeV, noncommutative
geometry predicts no new forces but a new uncertainty relation in
timespace at this energy scale. This uncertainty relation should shed
new light on quantum field theory.

\vskip .5cm

\noindent {\bf Acknowledgements:} As always, it is a pleasure to thank
Bruno Iochum for help and advice.

\vfil\eject

\begin{center}
\begin{tabular}{cccc}
\rxyz{0.7}{
,(10,-5);(5,-5)**\dir{-}?(.6)*\dir{>}
,(15,-10);(10,-10)**\dir{-}?(.6)*\dir{>}
,(10,-5);(10,-10)**\dir{-}?(.6)*\dir{>}
}   &
 $\;$$\;$ \rxyz{0.7}{
,(10,-5);(5,-5)**\dir{-}?(.6)*\dir{>}
,(15,-10);(10,-10)**\dir{-}?(.4)*\dir{<}
} &
 $\;$$\;$ \rxyz{0.7}{
,(10,-5);(5,-5)**\dir{-}?(.6)*\dir{>}
,(10,-15);(15,-15)**\dir{-}?(.6)*\dir{>}
}
 & $\;$$\;$
\rxyz{0.7}{
,(10,-5);(5,-5)**\dir{-}?(.6)*\dir{>}
,(5,-10);(15,-10)**\crv{(10,-13)}?(.6)*\dir{>}
}
 \\
\\ diag. 1 &  $\;$$\;$ diag. 2 &  $\;$$\;$ diag. 3 &  $\;$$\;$ diag. 4 \\
\rxyz{0.7}{
,(10,-5);(5,-5)**\dir{-}?(.6)*\dir{>}
,(10,-5);(10,-15)**\crv{(13,-10)}?(.6)*\dir{>}
}
 &  $\;$$\;$ \rxyz{0.7}{
,(10,-5);(15,-5)**\dir{-}?(.6)*\dir{>}
,(10,-15);(5,-15)**\dir{-}?(.6)*\dir{>}
}
 &  $\;$$\;$ \rxyz{0.7}{
,(5,-10);(5,-15)**\dir{-}?(.6)*\dir{>}
,(10,-15);(5,-15)**\dir{-}?(.6)*\dir{>}
}
 &  $\;$$\;$
\rxyz{0.7}{
,(15,-15);(10,-15)**\dir{-}?(.6)*\dir{>}
,(10,-5);(5,-5)**\dir{-}?(.6)*\dir{>}
,(10,-5);(15,-5)**\dir{-}?(.6)*\dir{>}
}
 \\
\\ diag. 5 &  $\;$$\;$ diag. 6 &  $\;$$\;$ diag. 7 &  $\;$$\;$ diag. 8 \\
\rxyz{0.7}{
,(15,-15);(10,-15)**\dir{-}?(.6)*\dir{>}
,(10,-5);(5,-5)**\dir{-}?(.6)*\dir{>}
,(5,-10);(5,-15)**\dir{-}?(.6)*\dir{>}
}
 &  $\;$$\;$ \rxyz{0.7}{
,(10,-5)*\cir(0.4,0){}*\frm{*}
,(10,-5);(5,-5)**\dir2{-}?(.6)*\dir2{>}
,(15,-15);(10,-15)**\dir{-}?(.6)*\dir{>}
}
 &  $\;$$\;$\rxyz{0.7}{
,(5,-5)*\cir(0.4,0){}*\frm{*}
,(10,-5);(5,-5)**\dir2{-}?(.6)*\dir2{>}
,(15,-15);(10,-15)**\dir{-}?(.6)*\dir{>}
} & 
$\;$$\;$
\rxyz{0.7}{
,(10,-5);(5,-5)**\dir{-}?(.6)*\dir{>}
,(5,-10);(5,-5)**\dir{-}?(.6)*\dir{>}
,(15,-15);(10,-15)**\dir{-}?(.6)*\dir{>}
} \\
\\ diag. 9 &  $\;$$\;$ diag. 10 &  $\;$$\;$ diag. 11 &  $\;$$\;$ diag. 12 \\
\rxyz{0.7}{
,(10,-5)*\cir(0.4,0){}*\frm{*}
,(10,-5);(5,-5)**\dir2{-}?(.6)*\dir2{>}
,(10,-15);(5,-15)**\dir{-}?(.6)*\dir{>}
}
 & $\;$$\;$ \rxyz{0.7}{
,(5,-5)*\cir(0.4,0){}*\frm{*}
,(10,-5);(5,-5)**\dir2{-}?(.6)*\dir2{>}
,(10,-15);(5,-15)**\dir{-}?(.6)*\dir{>}
}
 &$\;$$\;$ \rxyz{0.7}{
,(10,-5);(5,-5)**\dir{-}?(.6)*\dir{>}
,(5,-10);(5,-5)**\dir{-}?(.6)*\dir{>}
,(10,-15);(5,-15)**\dir{-}?(.6)*\dir{>}
}
 &
$\;$$\;$\rxyz{0.70}{
,(5,-15)*\cir(0.4,0){}*\frm{*}
,(10,-5);(5,-5)**\dir{-}?(.6)*\dir{>}
,(10,-15);(5,-15)**\dir2{-}?(.6)*\dir2{>}
}
\\
\\ diag. 13 &$\;$$\;$ diag. 14 & $\;$$\;$ diag. 15 & $\;$$\;$ diag. 16 \\
\rxyz{0.7}{
,(10,-15)*\cir(0.4,0){}*\frm{*}
,(10,-5);(5,-5)**\dir{-}?(.6)*\dir{>}
,(10,-15);(5,-15)**\dir2{-}?(.6)*\dir2{>}
}
 & $\;$$\;$ \rxyz{0.7}{
,(10,-5)*\cir(0.4,0){}*\frm{*}
,(10,-5);(5,-5)**\dir2{-}?(.6)*\dir2{>}
,(5,-15);(10,-15)**\dir{-}?(.6)*\dir{>}
}
 & $\;$$\;$ \rxyz{0.7}{
,(5,-5)*\cir(0.4,0){}*\frm{*}
,(10,-5);(5,-5)**\dir2{-}?(.6)*\dir2{>}
,(5,-15);(10,-15)**\dir{-}?(.6)*\dir{>}
}
 &
$\;$$\;$ \rxyz{0.7}{
,(10,-5);(5,-5)**\dir{-}?(.6)*\dir{>}
,(5,-10);(5,-5)**\dir{-}?(.6)*\dir{>}
,(5,-15);(10,-15)**\dir{-}?(.6)*\dir{>}
}

\\
\\ diag. 17 & $\;$$\;$ diag. 18 & $\;$$\;$ diag. 19 & $\;$$\;$ diag. 20 \\ 
\end{tabular}
\end{center}
\begin{center}
\begin{tabular}{cccc}
\rxyz{0.7}{
,(5,-15)*\cir(0.4,0){}*\frm{*}
,(10,-5);(5,-5)**\dir{-}?(.6)*\dir{>}
,(5,-15);(10,-15)**\dir2{-}?(.6)*\dir2{>}
}
 & $\;$$\;$ \rxyz{0.7}{
,(10,-15)*\cir(0.4,0){}*\frm{*}
,(10,-5);(5,-5)**\dir{-}?(.6)*\dir{>}
,(5,-15);(10,-15)**\dir2{-}?(.6)*\dir2{>}
}
 & $\;$$\;$ \rxyz{0.7}{
,(10,-5);(5,-5)**\dir{-}?(.6)*\dir{>}
,(10,-5);(10,-10)**\dir{-}?(.6)*\dir{>}
,(10,-15);(5,-15)**\dir{-}?(.6)*\dir{>}
}
 &
$\;$$\;$ \rxyz{0.7}{
,(10,-5);(5,-5)**\dir{-}?(.6)*\dir{>}
,(10,-5);(10,-10)**\dir{-}?(.6)*\dir{>}
,(5,-15);(10,-15)**\dir{-}?(.6)*\dir{>}
}
 \\
\\ diag. 21 & $\;$$\;$ diag. 22 & $\;$$\;$ diag. 23 & $\;$$\;$  diag. 24
\\
\\ 
\end{tabular}
\begin{tabular}{cc}
\rxyd{0.7}{ ,(15,-5);(5,-5)**\crv{(10,-8)}?(.6)*\dir{>}
,(25,-20);(5,-20)**\crv{(15,-24)}?(.6)*\dir{>} } &
$\;$$\;$$\;$$\;$$\;$
\rxyd{0.7}{ ,(15,-5);(5,-5)**\crv{(10,-8)}?(.6)*\dir{>}
,(25,-25);(15,-25)**\crv{(20,-22)}?(.6)*\dir{>}
,(5,-25);(5,-20)**\dir{-}?(.6)*\dir{>} }
\\
\\ diag. 25 & $\;$$\;$$\;$$\;$$\;$ diag. 26
\\
\\ 
\rxyddd{0.7}{
,(10,-5);(15,-5)**\dir{-}?(.6)*\dir{>}
,(20,-5);(25,-5)**\dir{-}?(.6)*\dir{>}
,(5,-30);(15,-30)**\crv{(10,-27)}?(.6)*\dir{>}
}
&
$\;$$\;$$\;$$\;$$\;$
\rxydd{0.7}{
,(5,-5);(5,-15)**\crv{(8,-10)}?(.6)*\dir{>}
,(20,-5);(15,-5)**\dir{-}?(.6)*\dir{>}
,(15,-25);(5,-25)**\crv{(10,-22)}?(.6)*\dir{>}
}

\\
\\ diag. 27 & $\;$$\;$$\;$$\;$$\;$ diag. 28
\\

\end{tabular}
\end{center}
\begin{center}
\begin{tabular}{cc}

\rxydd{0.7}{
,(5,-5);(15,-5)**\crv{(10,-8)}?(.6)*\dir{>}
,(20,-5);(15,-5)**\dir{-}?(.6)*\dir{>}
,(15,-25);(5,-25)**\crv{(10,-22)}?(.6)*\dir{>}
}
&
$\;$$\;$$\;$$\;$$\;$

\rxydt{0.7}{
,(25,-5)*\cir(0.4,0){}*\frm{*}
,(20,-5);(25,-5)**\dir2{-}?(.6)*\dir{>}
,(25,-15);(5,-15)**\crv{(15,-18)}?(.6)*\dir{>}
}
\\
\\ diag. 29 & $\;$$\;$$\;$$\;$$\;$ diag. 30
\\
\end{tabular}
\end{center}
\begin{center}
\begin{tabular}{cccc}
\rxyz{0.7}{
,(10,-5);(15,-5)**\dir{-}?(.6)*\dir{>}?(1)*\dir{(}
,(5,-15);(10,-15)**\dir{-}?(.6)*\dir{>}
}
&
\rxyz{0.7}{
,(10,-5);(5,-5)**\dir{-}?(.6)*\dir{>}?(1)*\dir{(}
,(15,-15);(10,-15)**\dir{-}?(.6)*\dir{>}
}
 & \rxyz{0.7}{
,(10,-5);(5,-5)**\dir{-}?(.6)*\dir{>}?(0)*\dir{)}
,(15,-15);(10,-15)**\dir{-}?(.6)*\dir{>}
}
    & \\
\\ diag. 31 & diag. 32 & diag 33 & \\

\rxyz{0.7}{
,(10,-5);(5,-5)**\dir{-}?(.6)*\dir{>}?(1)*\dir{(}
,(10,-15);(5,-15)**\dir{-}?(.6)*\dir{>}
}
 & \rxyz{0.7}{
,(10,-5);(5,-5)**\dir{-}?(.6)*\dir{>}?(0)*\dir{)}
,(10,-15);(5,-15)**\dir{-}?(.6)*\dir{>}
}
 & \rxyz{0.7}{
,(10,-5);(5,-5)**\dir{-}?(.6)*\dir{>}
,(10,-15);(5,-15)**\dir{-}?(.6)*\dir{>}?(1)*\dir{(}
}
 & \rxyz{0.7}{
,(10,-5);(5,-5)**\dir{-}?(.6)*\dir{>}
,(10,-15);(5,-15)**\dir{-}?(.6)*\dir{>}?(0)*\dir{||}
}
 \\
\\ diag. 34 & diag. 35 & diag. 36 & diag. 37 \\
\rxyz{0.7}{
,(10,-5);(5,-5)**\dir{-}?(.6)*\dir{>}?(1)*\dir{(}
,(5,-15);(10,-15)**\dir{-}?(.6)*\dir{>}
}
 & \rxyz{0.7}{
,(10,-5);(5,-5)**\dir{-}?(.6)*\dir{>}?(0)*\dir{)}
,(5,-15);(10,-15)**\dir{-}?(.6)*\dir{>}
}
 &\rxyz{0.7}{
,(10,-5);(5,-5)**\dir{-}?(.6)*\dir{>}
,(5,-15);(10,-15)**\dir{-}?(.6)*\dir{>}?(0)*\dir{)}
}
 & \rxyz{0.7}{
,(10,-5);(5,-5)**\dir{-}?(.6)*\dir{>}
,(5,-15);(10,-15)**\dir{-}?(.6)*\dir{>}?(1)*\dir{||}
}
 \\
\\ diag. 38 & diag. 39 & diag. 40 & diag. 41 \\
\\
\end{tabular}\linebreak\nopagebreak[4]
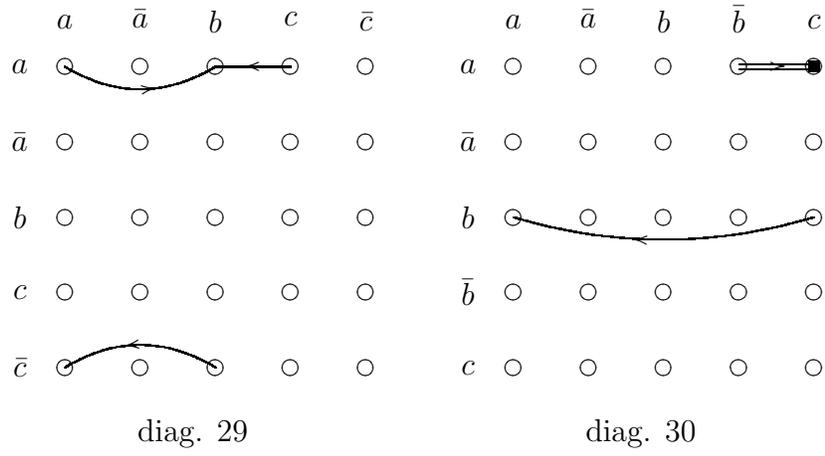
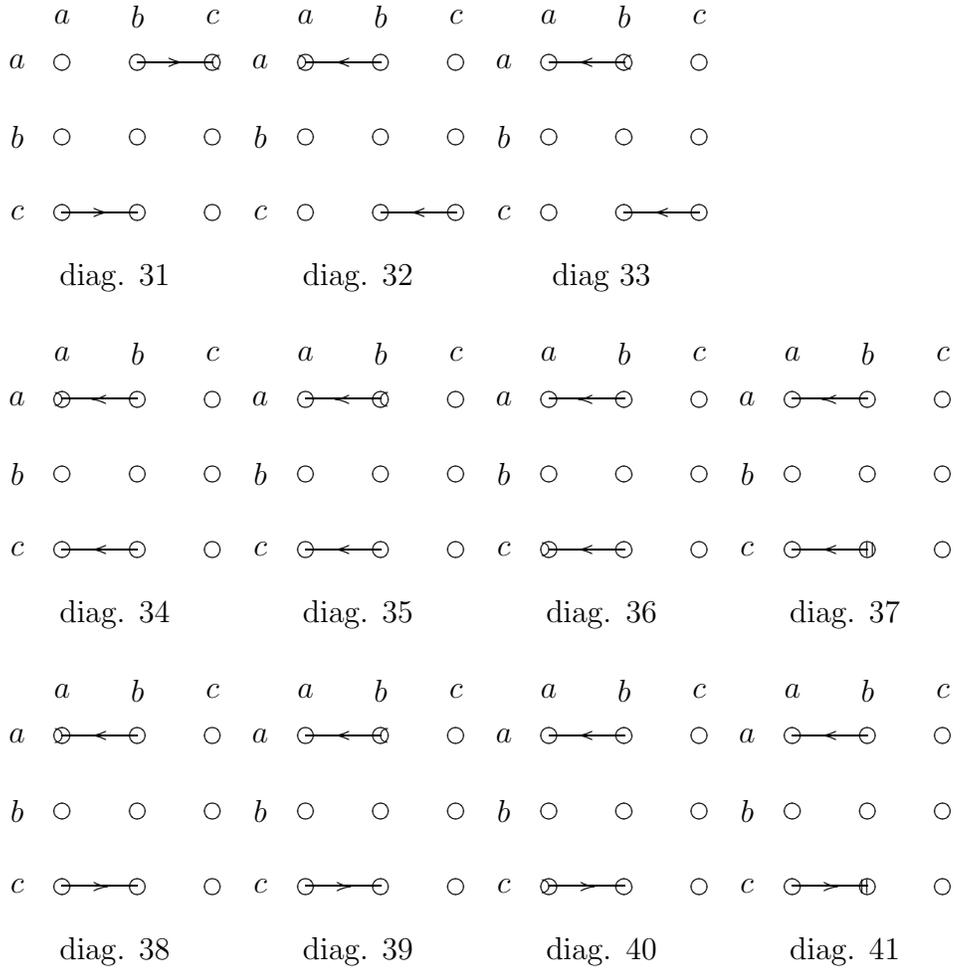
Figure 1: The 41 irreducible Krajewski diagrams for three summands
\end{center}

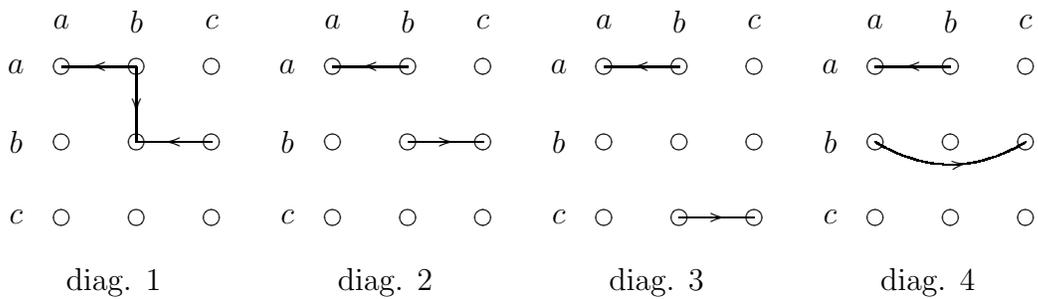
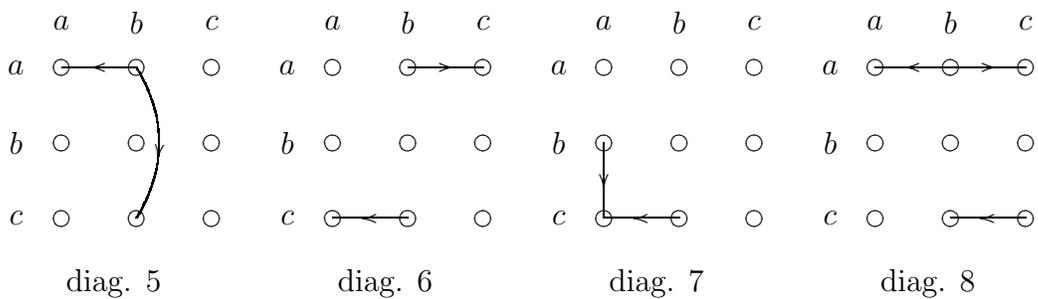
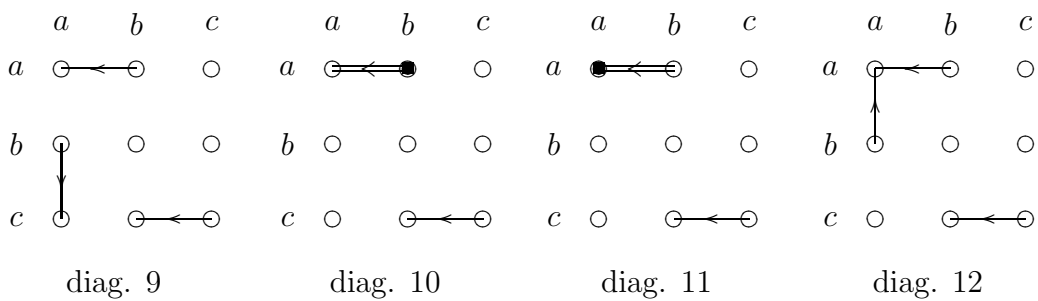
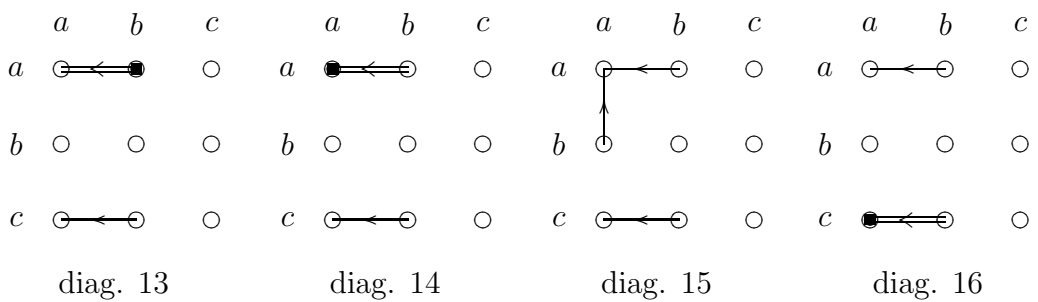
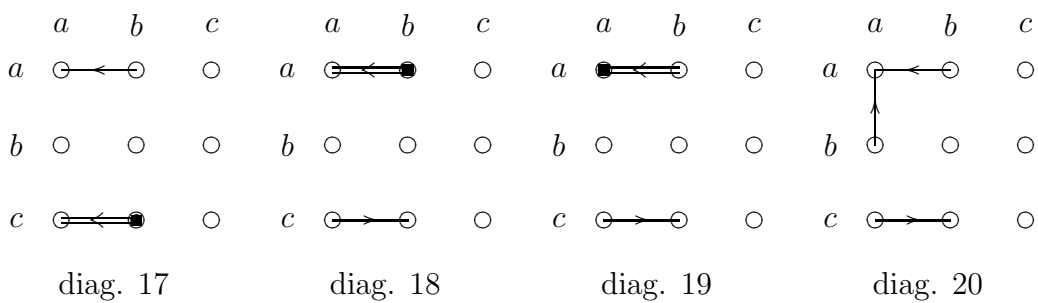
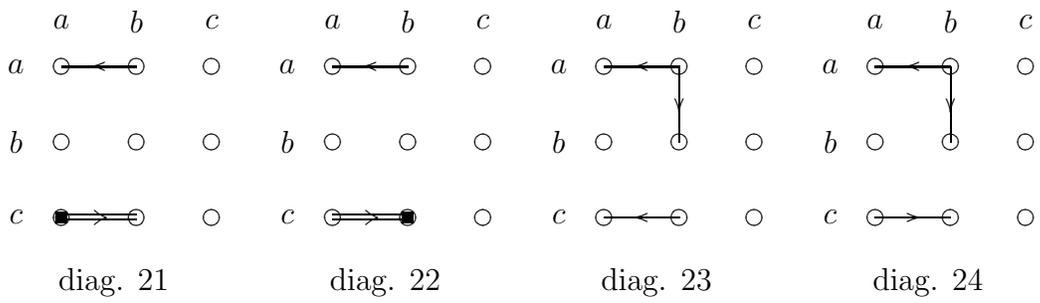
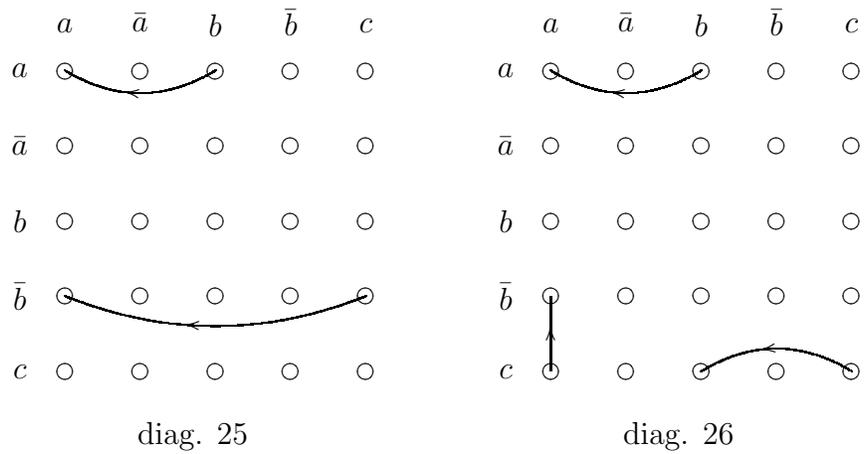
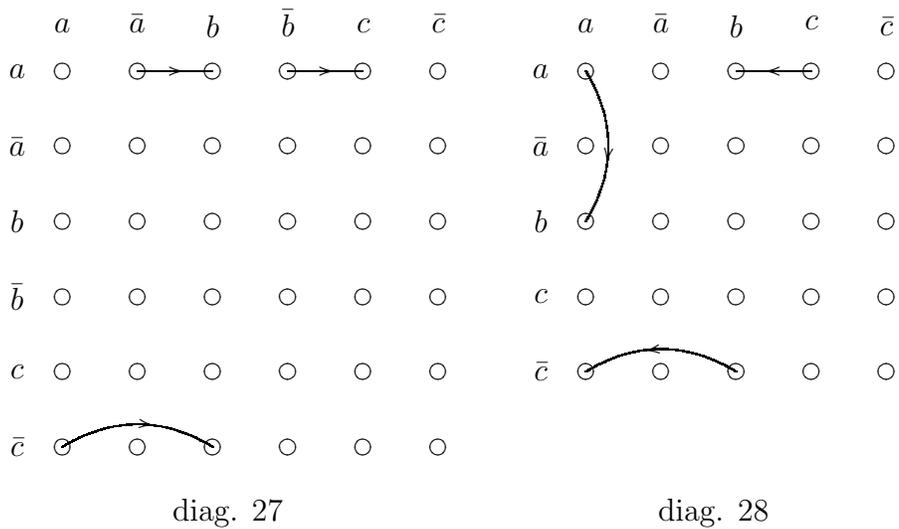
\end{document}